\documentclass{iau}

\usepackage{amsmath}
\usepackage{graphicx}
\usepackage{multirow}
\usepackage[percent]{overpic}

\begin{document}

\lefttitle{Testagrossa et al.}
\righttitle{Proceedings of the International Astronomical Union}

\jnlPage{1}{7}
\jnlDoiYr{2021}
\doival{10.1017/xxxxx}

\aopheadtitle{Proceedings IAU Symposium}
\editors{I. Liodakis, eds.}

\title{Accelerating SED Modeling of Astrophysical Objects Using Neural Networks}

\author{
  F. Testagrossa$^{1}$,
  G. Vasilopoulos$^{2}$,
  D. Karavola$^{2}$, 
  S.~I.~Stathopoulos$^{2}$,
  M.~Petropoulou$^{2}$, 
  C. Yuan$^{1}$,
  W. Winter$^{1}$
}

\affiliation{
  $^{1}$ Deutsches Elektronen-Synchrotron DESY, Platanenallee 6, 15738 Zeuthen, Germany\\
    $^{2}$ Department of Physics, National and Kapodistrian University of Athens, GR~15784, Zografos, Greece
}

\begin{abstract}
Interpreting the spectral energy distributions (SEDs) of astrophysical objects with physically motivated models is computationally expensive. These models require solving coupled differential equations in high-dimensional parameter spaces, making traditional fitting techniques such as Markov Chain Monte Carlo or nested sampling prohibitive. A key example is modeling non-thermal emission from blazar jets—relativistic outflows from supermassive black holes in Active Galactic Nuclei that are among the most powerful emitters in the Universe.
To address this challenge, we employ machine learning to accelerate SED evaluations, enabling efficient Bayesian inference. We generate a large sample of lepto-hadronic blazar emission models and train a neural network (NN) to predict the photon spectrum with strongly reduced run time while preserving accuracy. As a proof of concept, we present an NN-based tool for blazar SED modeling, laying the groundwork for future extensions and for providing an open-access resource for the astrophysics community.

\end{abstract}


\maketitle

\section{Introduction}
A fraction of active galactic nuclei (AGN) is known to launch relativistic and collimated plasma outflows, known as jets. A jetted AGN that is observed at a small angle with respect to the jet axis is called a \textit{blazar}~\citep{1995PASP..107..803U}.
Blazars are the most common extragalactic sources in the GeV gamma-ray sky \citep{2020ApJS..247...33A, 2022ApJS..260...53A}, but they are also known emitters of photons over many decades of energy, down to radio frequencies. Moreover, the variability timescale of their emission can range from minutes to months.

The rich phenomenology found in the spectral properties and in the time variability of blazar observations~\citep[see][for a review]{2019NewAR..8701541H} allows for different physical explanations of their spectral energy distribution (SED). Usually, the SED of blazars is interpreted in the framework of \textit{leptonic} models, where the radiation is emitted by a population of non-thermal relativistic leptons (electrons and positrons), interacting with the radiation fields in the magnetized jet.
Relativistic protons may also contribute to the observed high-energy blazar emission, but it is challenging to pin down the nature of radiating particles from electromagnetic observations alone \citep[for a review on blazar emission models, see][]{2020Galax...8...72C}. High-energy neutrinos are the smoking gun of relativistic hadrons in astrophysical sources, as they are produced through inelastic collisions of relativistic hadrons (protons and nuclei) with gas or photons. Therefore, the recent association of high-energy neutrinos with the blazar TXS~0506+056 \citep{2018Sci...361.1378I} has revived the interest in \textit{lepto-hadronic} models for blazars, where the electromagnetic radiation is produced by leptons and hadrons and neutrino emission is possible \citep[e.g.,][]{2018ApJ...864...84K, 2019NatAs...3...88G}. 

Fitting these models to the SED data allows to infer the physical parameters that better describe the properties of the jet environment and of the accelerated primary cosmic ray particles. However, connecting these inputs parameters to an output SED of radiation is a highly non-trivial task, as it requires to solve a network of coupled partial integro-differential equations, one for each particle species involved in the physical processes at place. Although some analytical approximations may be justifiable in certain conditions, a numerical approach is generally needed. This motivated the development of \textit{ad hoc} radiative codes~\citep[see][for a comparative presentation of the codes structure and
performance]{Cerruti24}, which evaluate numerically the model given a set of input parameters.
This model can then be used in a fitting routine for $\chi^2$-minimization to look for the best solution. However, this approach is not informative on the sensitivity of the results to the model parameters or on model degeneracies, which are particularly important in lepto-hadronic models. For this purpose, Markov Chain Monte Carlo (MCMC) or nested sampling fitting techniques are ideal, but their applicability has been limited due to the execution times of the underlying numerical codes used for the fitting, which can pile up to prohibitively long times. Addressing this computational bottleneck is central to increasing the pace of discovery in multi-messenger astrophysics. 

In this work, we train neural networks (NNs) on the simulations of the open-source codes {\tt AM$^3$} \citep{2024ApJS..275....4K} and {\tt LeHaMoC}~\citep{2024A&A...683A.225S} and use the NNs to predict an approximate photon spectrum.
The use of NNs as surrogate models significantly accelerates lepto-hadronic SED model evaluations, enabling efficient Bayesian inference. 

\section{Methodology}
In this section, we describe the dataset generation and the training of the NN that approximates the one-zone lepto-hadronic models of {\tt AM$^3$} and {\tt LeHaMoC}. In these models, the codes simulate the time evolution of primary electrons and protons, each injected with its own power-law spectrum $\frac{dN}{dE}\propto E^{-p}$, and of all the byproducts of the physical processes considered. The radiation zone is a relativistic blob moving inside the jet, and the numerical calculations are performed in the rest frame of the blob, where it is spherical and hosts a homogeneous magnetic field.

To produce the training data for the NN we compute the model on $2\times10^5$ sets of the ten input parameters, sampled in typical ranges for blazar modeling, shown in Table~\ref{tab:paramrange}.We include all the processes available in each code, which are listed in~\citep{Cerruti24}.   

The resulting photon spectra are then preprocessed and used as training data for the NN. In the design of the NN we focused on two architectures, deep neural networks (DNNs) and gated recurrent units (GRUs), exploring different combinations of hyperparameters. 
Our experiments showed that a GRU-based architecture gave the most accurate results for the lepto-hadronic model, confirming the findings of \cite{2024A&A...683A.185T} for leptonic models.

\begin{table}
\centering
\caption{Input parameter ranges used for the generation of training and validation datasets.\\
The parameters are intended in the blob rest frame. $U[\cdots]$ stands for a uniform distribution.} 
\resizebox{\textwidth}{!}{
\begin{tabular}{llc}
\hline
\textbf{Parameter} & \textbf{Symbol [unit]} & \textbf{Range}  \\
        \hline \hline
        Source radius &  $\log_{10} R$  [cm]   & $U[14, 18]$ \\
        Magnetic field strength & $\log_{10} B$  [G]   & $U[-2, 2]$ \\
        Electron injection luminosity & $\log_{10} L_e$ [erg s$^{-1}$] & $U[39,46]$ \\ 
        Electron minimum Lorentz factor & $\log_{10} \gamma_{e, \min}$ & $U[0.1, 4]$ \\
        Electron maximum Lorentz factor & $\log_{10} \gamma_{e, \max}$ & $U[\log_{10}\gamma_{e,\min} + 1, \min(9, \log_{10} \gamma_{e, \rm H})]$\textdagger \\
        Electron power-law slope & $p_{e}$ & $U[1,3]$ \\
        Proton injection luminosity & $\log_{10} L_p$ [erg s$^{-1}$] & $U[39,48]$ \\ 
        Proton minimum Lorentz factor & $\log_{10} \gamma_{p, \min}$ & $U[0.1, 4]$ \\
        Proton maximum Lorentz factor & $\log_{10} \gamma_{p, \max}$ & $U[\log_{10}\gamma_{p,\min} + 1, \min(9, \log_{10} \gamma_{p, \rm H})]$\textdagger  \\
        Proton power-law slope & $p_{p}$ & $U[1,3]$ \\
        \hline
\end{tabular}
}
\textdagger \, The symbol $\gamma_{i, \rm H}$ indicates the Hillas-limited Lorentz factor of particle species $i$, which is defined as $\gamma_{i, \rm H}= eBR / m_i c^2$~\citep{1984ARA&A..22..425H}. 
\label{tab:paramrange}  
\end{table}

\begin{figure}[!ht]
\centering
\begin{overpic}[width=\columnwidth]{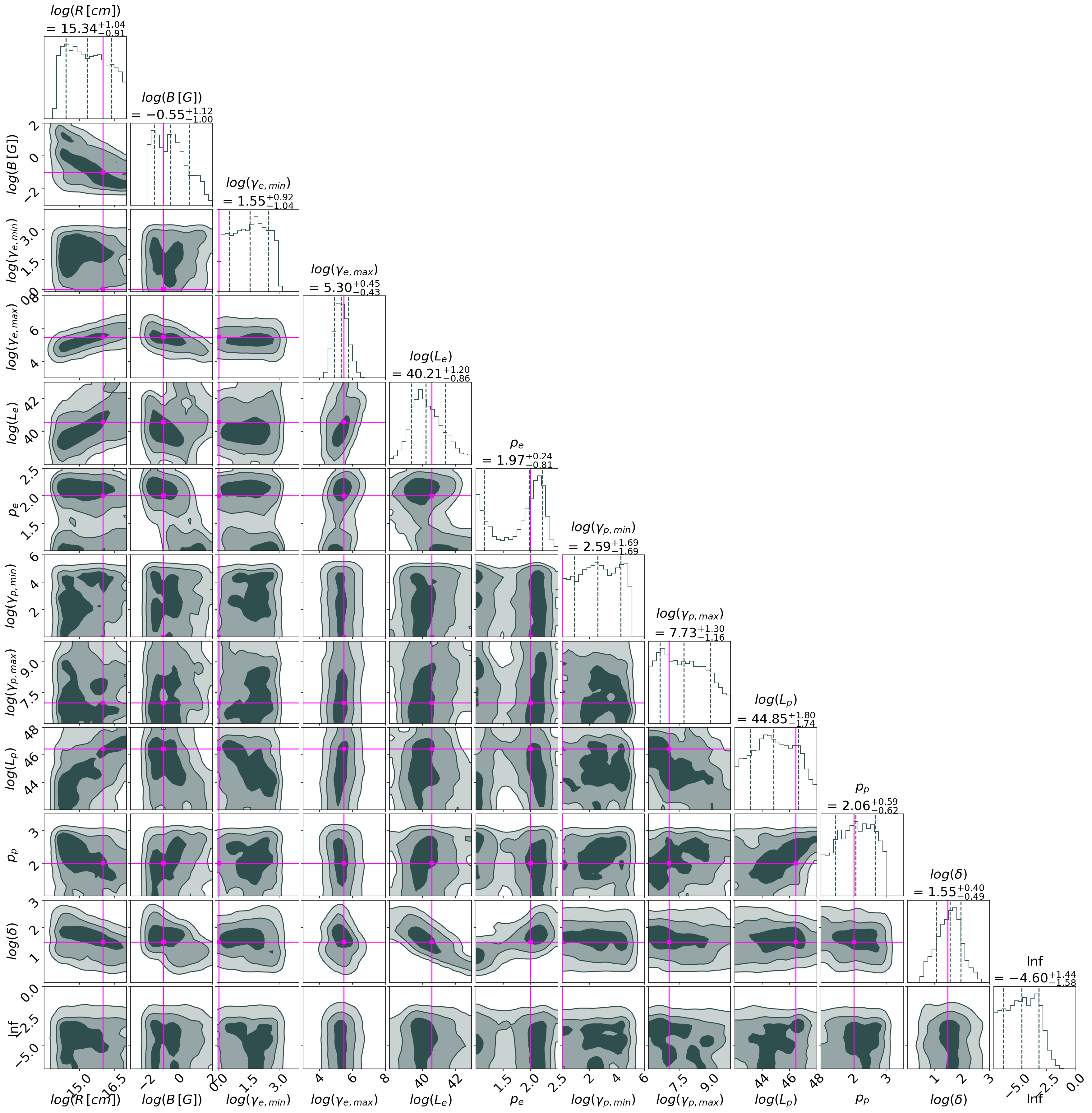}
    \put(43,61){\includegraphics[width=0.57\columnwidth]{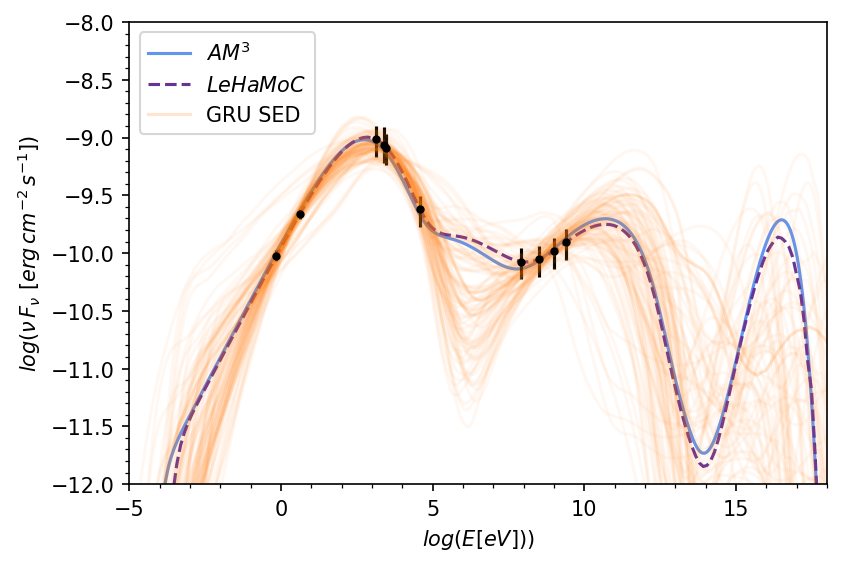}}
\end{overpic}
\caption{Corner plot of the posterior distribution of model parameters computed by the NN fit to the simulated data. In the upper inlay panel we display the SED with the simulated dataset, model estimated based on numerical codes and 100 predictions of the NN model.}
\label{fig:corner_with_sed}
\end{figure}
 
\section{Results}
Here, we will report the results for the NN trained on the {\tt AM$^3$} dataset, as the training on the {\tt LeHaMoC} dataset is still ongoing. A comprehensive comparative analysis of the two trained NNs will be presented in a forthcoming publication \citep{Testagrossa26}, together with the final architecture of the NN.

To evaluate the performance of the trained NN, we performed a blind fit to a simulated blazar SED. This was generated with {\tt LeHaMoC} using the parameters of the LeHa model described in \citet{Cerruti24} (see their Table~3). From this SED, we randomly sampled two points in the range $10^{14}–10^{15}\,\text{Hz}$, four points between $10^{17.3}–10^{19}\,\text{Hz}$, and four additional points within $10^{22}–10^{25}\,\text{Hz}$, corresponding to the typical observational windows of optical, X-ray, and GeV gamma-ray instruments. A 10\% and 30\% uncertainty was then added to the optical and higher energy flux measurements, respectively. We then fitted the data with the NN as a surrogate model, employing likelihood minimization with {\tt emcee} with the addition of an excess variance parameter $lnf$
\citep[for methodology see;][]{2023MNRAS.520..281K,2024A&A...683A.225S}.
Since the NN computes the photon spectrum in the jet frame, its output has to be boosted to the observer frame:  the energy must be multiplied by a Doppler factor $\delta$ and the intensity by $\delta^4$. The boost is an analytical operation that can be performed after the NN evaluation, so we implemented it in the routine by considering $\delta$ as an additional free parameter of the fit. The MCMC fit used 48 walkers and 20,000 steps, with convergence achieved in less than 2,000 steps based on the estimated autocorrelation time. The fit was completed in one hour on a standard Mac laptop. For comparison, computing one million models with the {\tt AM$^3$} code would have required approximately 10k to 20k hours on the same machine.

The simulated SED and the corresponding corner plot of the posterior distributions from the fit are shown in Fig.~\ref{fig:corner_with_sed}. In the SED panel, we also display the models computed with the {\tt AM$^3$} and {\tt LeHaMoC} codes, along with 100 random realizations drawn from the posterior distribution and calculated with the NN. The two numerical codes produce very similar results (maximum difference is at the level of 20\%). Moreover, the NN-based SED model reproduces the simulated data points accurately, and the parameters used to generate the dataset are generally recovered by the fit, lying close to the median values of the posterior distribution, with only few exceptions. The fit to the simulated SED also highlights the importance of coverage in specific energy bands. In particular, the absence of observations in the MeV and TeV ranges introduces degeneracies in the fitted models. Consequently, the model’s flexibility allows a wide range of SED shapes to be consistent with the data, resulting in broad posterior distributions. 

\section{Conclusions}
We have successfully trained a lepto-hadronic GRU-based NN for blazar emission based on the open-source code {\tt AM$^3$} and demonstrated its performance through MCMC SED fitting. 
In the future, we will add more features to the models used in the training, including neutrino spectra, and then we will ultimately provide the trained NN to the community.

\vspace{-0.1in}
\section*{Acknowledgments}
\small 
We thank Margot Boughelilba for useful comments on this
manuscript. M.P., D.K., S.I.S. acknowledge support from the Hellenic Foundation for Research and Innovation (H.F.R.I.) under the
“2nd call for H.F.R.I. Research Projects to support Faculty members and Researchers” through the project UNTRAPHOB
(Project ID 3013). G.V. acknowledges support from the H.F.R.I. through the project ASTRAPE (Project ID 7802).
The authors also acknowledge support from the “Program for the Promotion of Exchanges and Scientific
Collaboration between Greece and Germany IKYDA–DAAD” 2024 (IKY project ID 309; DAAD project ID 57729829).



\end{document}